\documentclass[rmp,floatfix,twocolumn,tightenlines]{revtex4} 
\makeatletter
\bibpunct{(}{)}{,}{n}{,}{,}
\usepackage[pdftex]{graphicx}
\usepackage{amsmath}

\begin{document} 

\title{ Beyond similarity: A network approach for identifying and delimiting biogeographical regions }

\author{Daril A. Vilhena}
\affiliation{Department of Biology, University of Washington, Seattle, WA 98195-1800. E-mail: daril@uw.edu}
\author{Alexandre Antonelli}
\affiliation{ Department of Biological and Environmental Sciences, University of Gothenburg \& Gothenburg Botanical Garden, G\"{o}teborg, Sweden. E-mail: alexandre.antonelli@bioenv.gu.se}

\begin{abstract}

Biogeographical regions (geographically distinct assemblages of species and communities) constitute a cornerstone for ecology, biogeography, evolution and conservation biology. Species turnover measures are often used to quantify biodiversity patterns, but algorithms based on similarity and clustering are highly sensitive to common biases and intricacies of species distribution data. Here we apply a community detection approach from network theory that incorporates complex, higher order presence-absence patterns. We demonstrate the performance of the method by applying it to all amphibian species in the world (c. 6,100 species), all vascular plant species of the USA (c. 17,600), and a hypothetical dataset containing a zone of biotic transition. In comparison with current methods, our approach tackles the challenges posed by transition zones and succeeds in identifying a larger number of commonly recognised biogeographical regions. This method constitutes an important advance towards objective, data derived identification and delimitation of the world's biogeographical regions.

\end{abstract}

\date{\today}
\maketitle

\section{INTRODUCTION}

Considerable attention has been devoted to develop methods that can confidently assign individuals to populations \cite{baudouin2004analytical,hansen2001assigning}, and then group those populations into phylogenetic entities that deserve the status of species or evolutionary units \cite{de2007species}. How species then co-exist and co-interact to form clusters at higher levels, of similar taxonomic and eco-physiological characteristics, is much less understood. This is surprising, considering that already by the 19\textsuperscript{th} century prominent naturalists such as Humboldt and Bonpland \cite{von1807essai}, de Candolle \cite{de1820essai}, Prichard \cite{prichard1826researches}, Sclater \cite{sclater1858general} and Wallace \cite{wallace1876geographical} had all realised that the world's biota is divided into a number of more or less distinct units. 

The recognition and use of biological regions, or bioregions, offers several advantages as compared to studying individual species or communities, and has therefore gained in popularity in the last years in both terrestrial and aquatic systems \cite{spalding2007marine,gonzalez2013biogeographical,abell2008freshwater,olson2001terrestrial}. A bioregion based approach in macroecology and evolution can be used to assess to what extent lineages are able to cross major eco-physiological barriers over evolutionary time, i.e. their degree of niche conservatism in a broad sense \cite{wiens2010niche,crisp2009phylogenetic}. Evidence is growing that different bioregions will be affected differently by climate change \cite{salazar2007climate,knapp2001variation}, so understanding their origins and evolution \cite{pennington2006insights,crisp2006biome} may provide further indications of their expected resilience to future climate changes \cite{condamine2013macroevolutionary}. Bioregions may also be used as operational units in ancestral reconstruction analyses, aimed at inferring key biogeographic processes (dispersal, vicariance, speciation and extinction) for particular lineages \cite{silvestro2011bayesian}. Finally, a cross-taxonomic approach based on bioregions also offers important advantages in conservation biology as compared to focus on single taxons, not least in species rich areas such as seasonally dry tropical forests \cite{toby2000neotropical,sarkinen2011forgotten}. In such areas, conservational efforts may be better targeted towards protecting remaining patches of threatened bioregions rather than focusing on particular species. In this sense, bioregions may be considered analogous to Biodiversity Hotspots, a concept based on species richness, endemicity and threat, which has received enormous attention in ecology, biogeography and conservation in the last decades \cite{mittermeier2011global}. 

Many studies take for granted the identity and delimitation of biogeographical regions around the world. Yet, there is little agreement on how to best classify and name such regions, with several conceptually related terms being used, often interchangeably \cite{allaby2010dictionary}\cite{morrone2014biotas}. These include biomes, ecoregions, realms, provinces, zoo/phyto-geographic regions, ecosystems, ecozones,  chorotypes, dominions, areas of endemism, concrete biota, chronofauna, nuclear area, horofauna, cenocron, phytocorion, generalised track, biogeographical/taxonomic/species assemblage, and domains. Regionalisation concepts vary among disciplines (e.g. between zoology and botany) and regions, with e.g. Africa having a generally accepted system for plants \cite{white1993aetfat}, whereas South America lacks a unified, congruent floristic classification \cite{sarkinen2011forgotten}. Moreover, different names may apply to the same unit; examples in South America include the Cerrado vs the Brazilian savanna, and the P{\'a}ramo vs high-altitude Andean grasslands \cite[e.g.,][]{hughes2013neotropical}. 

One common feature in most classification systems is an internally implied hierarchy. This is for instance evident in the terrestrial classification system of Olson \textit{et al}.\cite{olson2001terrestrial}, which is the one adopted by the World Wide Fund for Nature (WWF) and recognises 8 realms, nesting 14 biomes which in turn contain 867 ecoregions. In that scheme, ecoregions are defined as \textit{"relatively large units of land containing a distinct assemblage of natural communities and species, with
boundaries that approximate the original extent of natural communities prior to major land-use change"} and reflecting \textit{"distributions of a broad range of fauna and flora across the entire planet"}. This and other classification systems widely used in biogeography (e.g., \cite{wallace1876geographical}) include a key taxonomic component, which contrasts with purely abiotic approaches such as in the K{\"o}ppen-Geiger Climate Classification \cite{koppen1900versuch}, which in its latest update \cite{kottek2006world} is based solely on ranges of temperature, precipitation, and their distribution over the year.

\begin{figure*}[!ht]
\begin{center}
\includegraphics[scale=1]{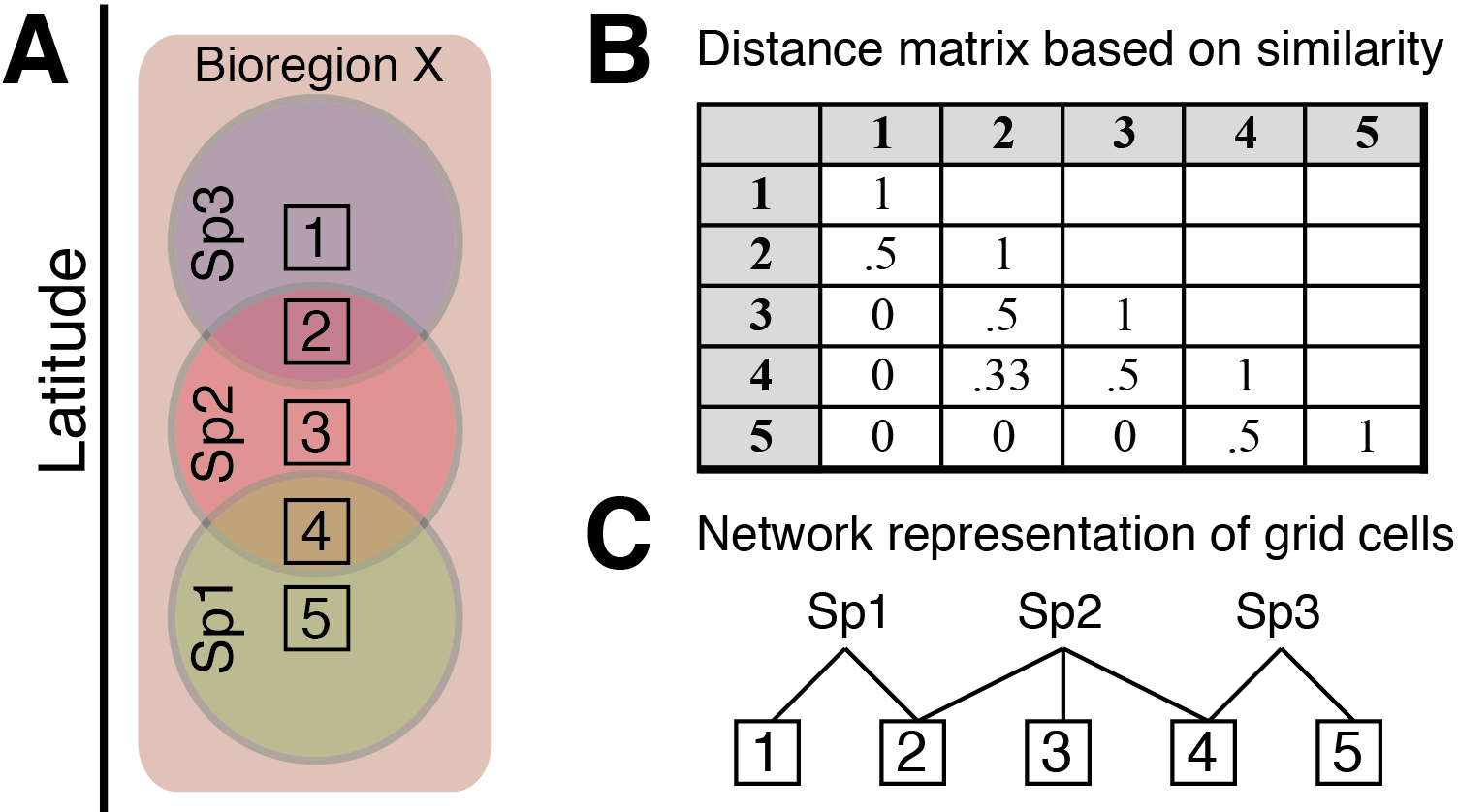}
\end{center}
\caption{
{\bf Comparison between similarity based clustering and the network method developed here}. A) Three species (Sp1, Sp2, Sp3) occur in the generally recognised Bioregion X, which spans a large latitudinal gradient. Species diversity is measured from five grid cells (numbered 1--5). Note that there is little geographic overlap between the species ranges, represented by circles. B) Diversity similarity (set measures) between grid cells, which computes the similarity in number of shared species (the Jaccard index is shown here). Note that the distance between grid cell 1 and 5 is zero, since they do not share any species. C) In the network method, connectivity between grid cells is established through the species they contain. In this case, grid cells 1 and 5 are ``connected" by a single step through one species (Sp2), which does not occur in either cell but occurs in other cells (2 and 4) occupied by species that also occur in cells 1 and 5 (Sp1 and Sp3, respectively). See text for more details.
}
\label{Figure 1}
\end{figure*}

Perhaps more importantly than the lack of consensus in terminology and classification system used for biogeographical regions -- which is to some extent more of a semantic issue rather than a true biological problem \cite{morrone2014biotas} -- there remains controversy on how to best identify and delimit these regions regardless of hierarchical status. In the last decades, deductive approaches have been replaced by more analytical, transparent and reproducible methods \cite{hagmeier1964numerical,kreft2010framework,holt2013update}. Bioregionalisation based on species distribution data needs to deal with particular challenges such as biased taxonomic sampling. Even so, it has been shown to outperform even high resolution remote sensing techniques that rely on structural differences in vegetation \cite{sarkinen2011forgotten} and may be more sensitive to human-mediated effects on the landscape, such as changes in land use and land cover (e.g. clearing, plantations, irrigation, drainage, urbanisation). 

The detection of bioregions is impacted by how we choose to quantify biogeographic structure, which up to now has been chiefly a variety of species turnover measures based theoretically on beta diversity \cite{koleff2003measuring, hagmeier1964numerical, kreft2010framework}. Species turnover, as measured by set-based similarity measures such as the Jaccard \cite{jaccard1901etude}, S{\o}renson \cite{dice1945measures}, and $\beta$-similarity \cite{koleff2003measuring}\cite{simpson1943mammals}, quantifies the relationship of one region to another, typically by dividing the number of shared species between two regions by some measure of the total species in both regions \cite{tuomisto2010diversity}. 

Despite their widespread use, species turnover measures can miss intricacies of distributional data that are relevant for bioregion detection. \textit{First}, species turnover tends to increase with greater geographic distance from a source, bringing into question whether geography or biogeography drives the pattern \cite{mccoy1987some}. \textit{Second}, for small spatial scales turnover can overestimate disparity due to competitive exclusion, spatial clustering, and environmental gradients \cite{vellend2001commonly}. Although this problem can be reduced with large plot sizes, the problem is expected to persist even for large spatial scales. Furthermore, competitive exclusion can create geographic boundaries between species that cohabit the same bioregion. \textit{Third}, some generally recognised bioregions span many degrees of latitude, such as the North American Rocky Mountains and the American Great Plains, and may contain climatic and environmental heterogeneities that can cause narrowly distributed taxa to occupy non-overlapping fractions of the same bioregion (Fig. 1). \textit{Fourth}, differences in taxonomic sampling are expected to inflate turnover. For example, taxonomic standards may differ within bioregions for rare species. For deep time studies, marine fossil assemblages may for instance not co-preserve aragonitic and calcitic shells. These processes collectively bias turnover measures, because the number of shared species cannot always be trusted as good gauge of bioregion identification. 

Here we present a data driven approach that uses associational networks to minimise the problems described above and to extract more community level information from species occurrence data. We show that this method can be used to successfully detect bioregion level regions in one hypothetical and two well validated empirical datasets: all amphibians at a global level, and all vascular plants in the United States of America. The empirical datasets provide contrasting examples of how biodiversity data is currently available: they are aggregated at different scales (global and national), grain sizes (two degree grid cells vs US counties), and were constructed under different sampling methodologies. Our results are strikingly congruent with opinion generated bioregion delimitations, indicating that the network method developed here holds the potential to improve the identification and delimitation of the world's bioregions. 

\section{Results} 

\subsection{Amphibians} 

By abstracting species distributions as a network, we were able to incorporate complex presence-absence relationships into bioregion delineation, whereas with similarity measures only the percentage of shared species can be included. In the occurrence network, bioregions appear as groups of localities and taxa that are highly interconnected. Figure 2B shows a visualisation of the network of all native amphibian species. In this network, the broad spatial separations of clusters represent the realms, while the bioregions are coloured differently within each larger cluster. The links that cross between realms correspond to the relatively few widespread species that inhabit multiple bioregions on multiple realms and continents. 

\begin{figure}[!ht]
\begin{center}
\includegraphics[scale=.2]{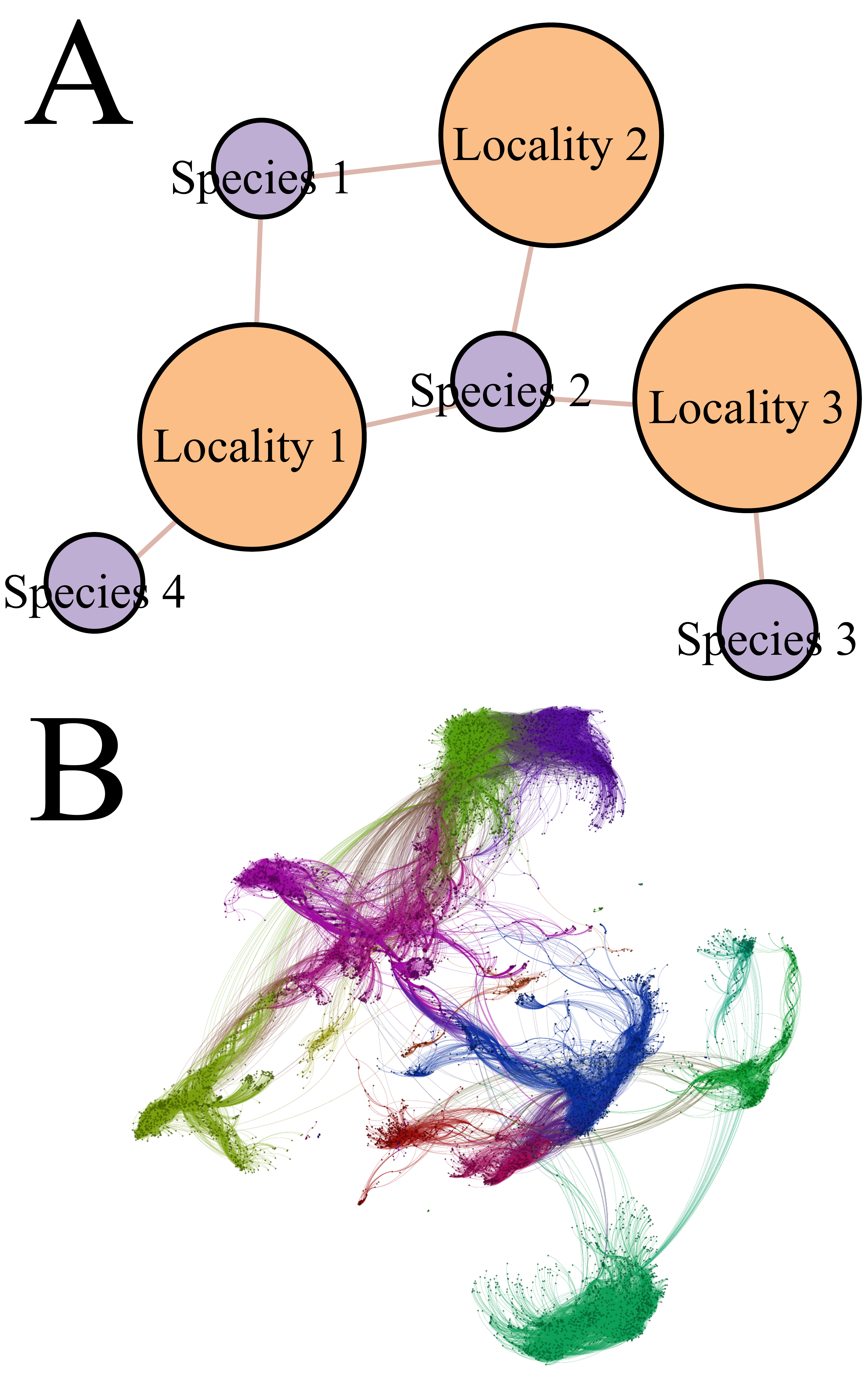}
\end{center}
\caption{
{\bf Bipartite occurrence network}. (a) Schematic representation showing the different classes of network connectivity that can be formed. Species 1 and 2 jointly occur in Locality 1 and 2, which creates a 4-path that loops, while Species 3 and 4 share a 4-path that does not loop, revealing that the species range of an intermediary species (Species 2)``connects" the two. (b) A visualisation of the amphibian network analysed here (N=6,100 species). The geographic ranges of widespread species act as highways between biogeographic realms, creating links between clusters. Node positions determined by the Force Atlas algorithm in the Gephi package \cite{bastian2009gephi}. 
}
\label{Figure 2}
\end{figure}

Our analysis identified 10 zoogeographic realms and 55 bioregion level regions as the optimal representation of the full amphibian dataset (Figure 3A). This differs from the approach in Holt \textit{et al}. \cite{holt2013update} using a species turnover measure based on $p\beta$sim, which identified 19 bioregions as optimal. To illustrate how well range limits reflect bioregion structure, we coloured geographic ranges by the region they were assigned to (Figure 3B). These results differ not only quantitatively but also qualitatively, showing differences in major biogeographic boundaries. For instance, the network method is able to detect both Wallace's and Weber's line as boundaries, but Weber's line emerges as the major boundary between the Oceanian and Oriental faunas \cite[see e.g.][for a detailed account on the different lines proposed to separate these faunas]{simpson1977too}. Moreover, our approach reveals Sulawesi and the islands between Wallace's and Weber's line as distinct subregions of the Oriental realm.

\begin{figure*}[!ht]
\begin{center}
\includegraphics[scale=.7]{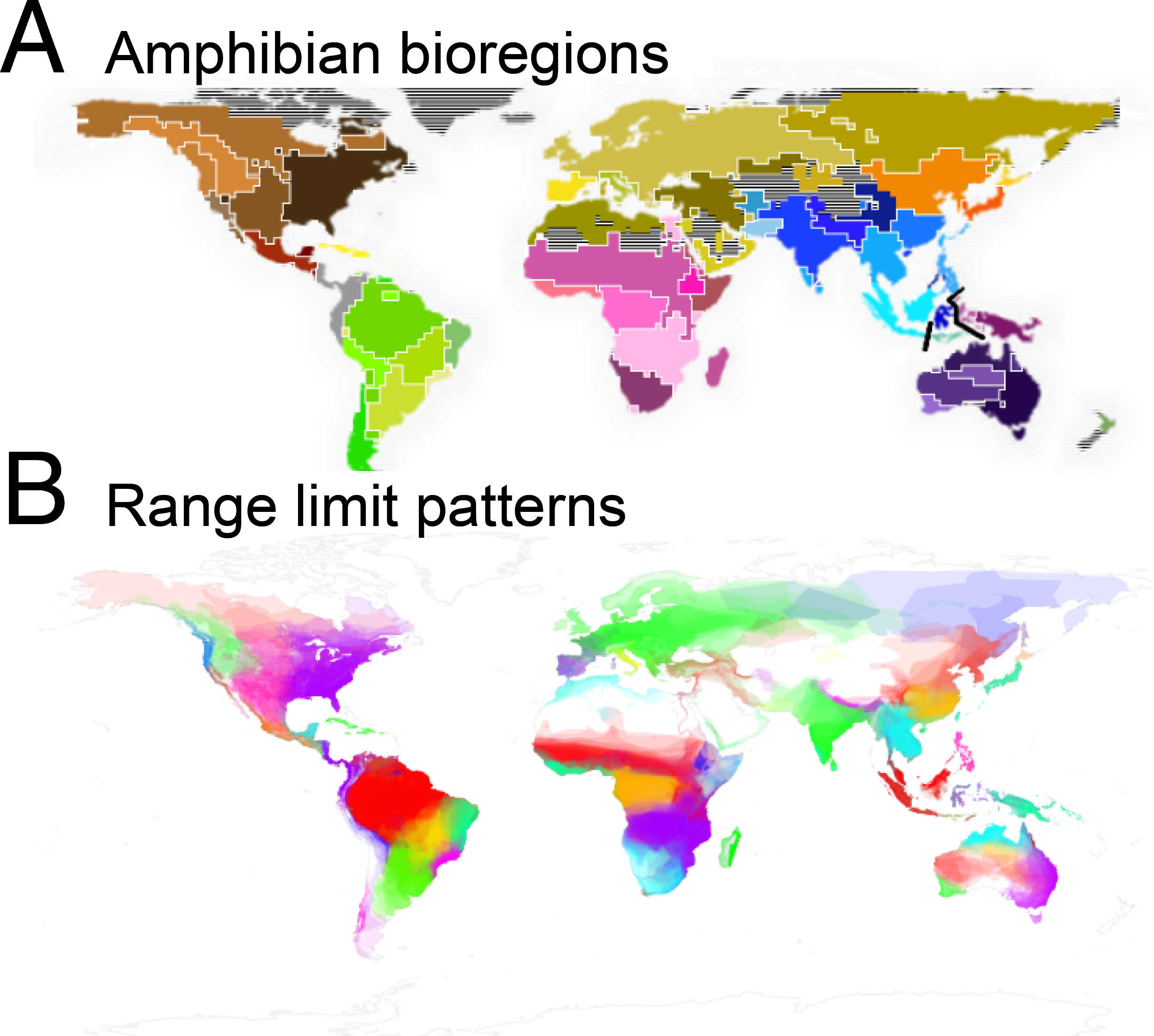}
\end{center}
\caption{
{\bf Results from the network analyses for the world's amphibians (N=6,100 species)}. A) Amphibian Biogeographical regions of the world determined from geographic range data. Similar colours indicate membership to a higher level realm. The analysis used a resolution of two degree grid cells. B) Species range limits coloured by region. Geographically close and neighbouring regions were given contrasting colours to highlight boundaries and boundary mixing. Each geographic range polygon was plotted with a low opacity (0.1), from largest to smallest, so that regions with more species appear brighter. 
}
\label{Figure 3}
\end{figure*}

\subsection{Vascular plants}

\textbf{Similarity approach.} Applying the similarity approach to our three USDA datasets of native plants, the number of clusters selected as optimal was 11 for all plants, 22 for trees, and 14 for non-trees. The resulting optimal partition of counties for all datasets (Fig. 4, middle column) reveals little biogeographic structure. For all native plants, the boundary between the two largest clusters approximates the boundary between the American Great Plains and Eastern Temperate Forests, but it is dominated by rigid state boundaries and fails to distinguish, for example, the Everglades in southern Florida, the Pacific Coast, and the Rocky Mountains. The tree dataset separates the Everglades from the rest of the United States, and the non-tree dataset mimics the major boundaries in the all plant dataset but contains more clusters that are also US states.

\begin{figure*}[!ht]
\begin{center}
\includegraphics[scale=.2]{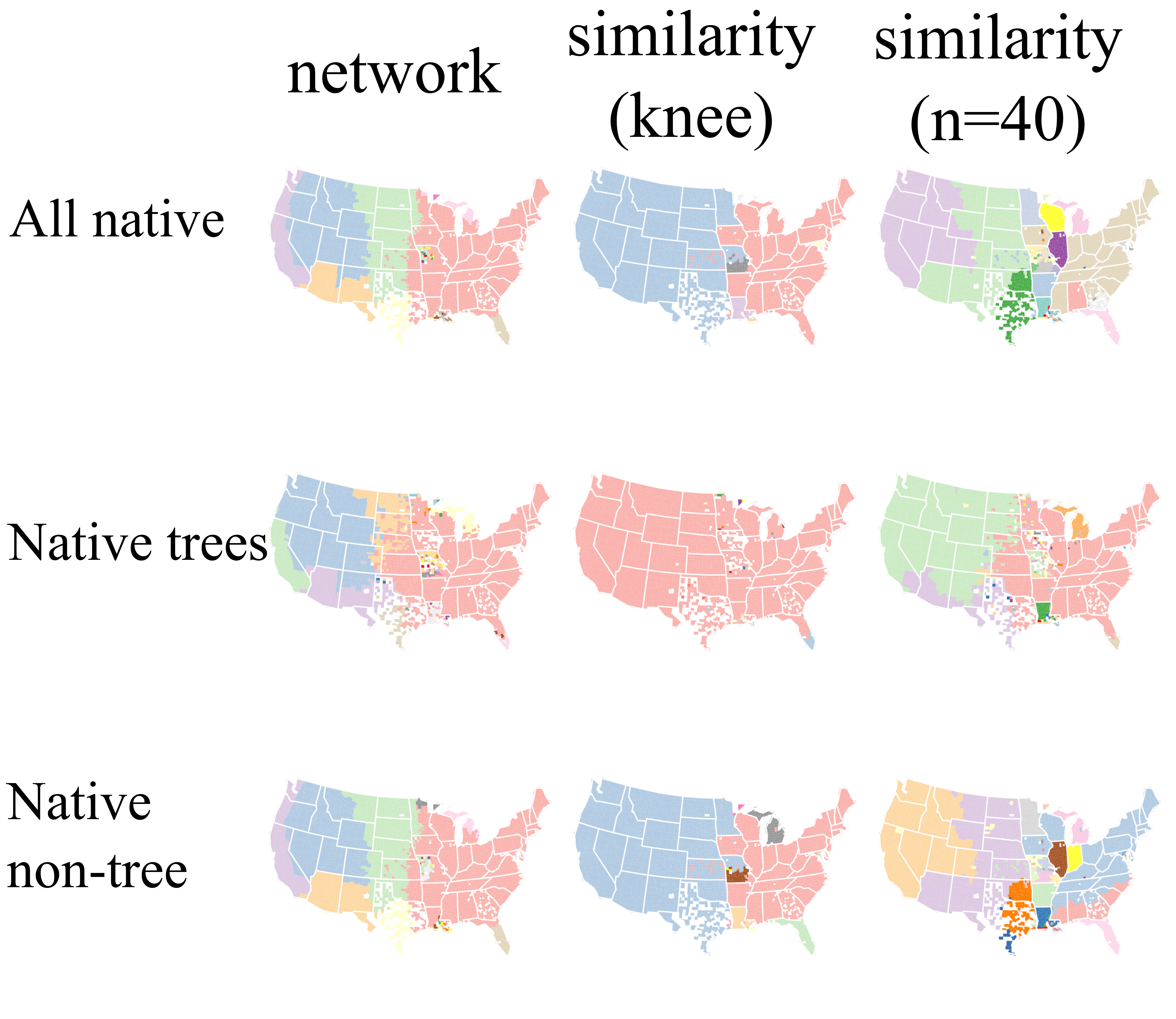}
\end{center}
\caption{
{\bf United States plant bioregions (N=17,600 taxa)}. Demarcations of american bioregions for three subsets of the USDA plant database: all native plants, native trees, and non-native trees. The left column was determined by the map equation (optimal number of clusters shown), while the middle and right columns were determined by a similarity approach (optimal number of clusters and an arbitrarily finer scale delineation, respectively). Overall, the network approach captures with broad brushstrokes the general patterns of the United States biomes. Although state level biases are apparent from both methodologies, they are strikingly more recurrent in the similarity approach. 
}
\label{Figure 4}
\end{figure*}

To explore whether the similarity approach could be arbitrarily forced to unveil deeper structure, we also chose to visualise the partitions with 40 clusters selected, although this delineation is not optimal (Fig. 4, right column). Some biogeographic structure becomes apparent at this level -- the American Great Plains is cleanly separated from the American West, although this bioregion unrealistically stretches into the American Southwest desert. The reconstruction based on these 40 clusters is also plagued by a number of boundaries coincident with US state boundaries in the American midwest. In the tree-level data, the Great Plains division becomes apparent, as well as a clean separation of the Southwest desert from the American West. In the non-tree dataset, a latitudinal boundary is evident in the Eastern Temperate Forests bioregion, but also contains ample state level biases. 

\textbf{Network approach.} We generated a network dataset from the USDA plant data, with county nodes connected to species nodes if the species was identified as natively present in that county. From these data, we clustered the map equation. A pilot analysis revealed little hierarchical structure in the data, so we opted to use a two-level implementation of the map equation, which produces $k$ clusters instead of hierarchically nested groups of clusters \cite{rosvall2008maps}. The apparent lack of hierarchy in the data is likely an issue of large grain and low scale (counties within a single country). Higher resolution data, such as a database produced from geographic coordinates, might produce greater subdivision and not require this additional implementation. Applied to our three USDA datasets of native plants, the number of clusters selected as optimal by the map equation was $25$ for all plants, $19$ for trees, and $16$ for non-trees. Because the algorithm that seeks the best partition is stochastic, we ran it 1000 times and selected the partition that minimised the scoring function in the map equation. 

Broad similarities are evident across the clustering results for all native plants, trees, and non-trees (Fig. 4, left column). There are, however, a number of differences. For instance, the Everglades are only evident from the tree-only dataset. The West Coast forms a separate bioregion under the analysis of all plants as well as all non-trees, but the Pacific Northwest is omitted from this bioregion when only trees are considered. In the American midwest, the American Great Plains appear much smaller when only trees are considered. These differences may reflect intrinsic biological differences among the datasets analysed (e.g. differences in ecological niche conservatism, edaphic adaptations, dispersal ability), but sampling issues are also apparent. For instance, the southern deserts of Arizona and surrounding area follow some rigid state boundaries, suggesting that large county sizes in the area obscure finer demarcation. State level biases are also evident in Lousiana for the native tree data, but not for the other two datasets. 

\subsection{Hypothetical dataset}

Using Simpson's similarity index and UPGMA on the hypothetical dataset of Kreft \& Jetz \cite{holt2013updatereply}, the transition zone is engulfed by the Northern realm for a choice of two clusters, and it is a distinct cluster if three clusters are chosen (Fig. 5B). The data are symmetric, so if the matrix rows are swapped the transition zone is engulfed by the Southern realm. 

Applying the network method to the same data results in an optimal partition of four clusters: one contains all of the Southern fauna and grid cells 1-14, one contains all of the Northern fauna and grid cells 17-30, while grid cells 15 and 16 each form their own cluster (Fig. 5C). This partition is slightly preferred over a two cluster solution, which evenly cuts the data into two biogeographic zones. This example reveals the benefit of clustering both species and grid cells together, as opposed to clustering grid cells with distances proportional to the number of shared species -- grid cells 15 and 16 can easily be identified as transition zones because no species are clustered with them (Fig. 5C).

\begin{figure*}[!ht]
\begin{center}
\includegraphics[scale=.18]{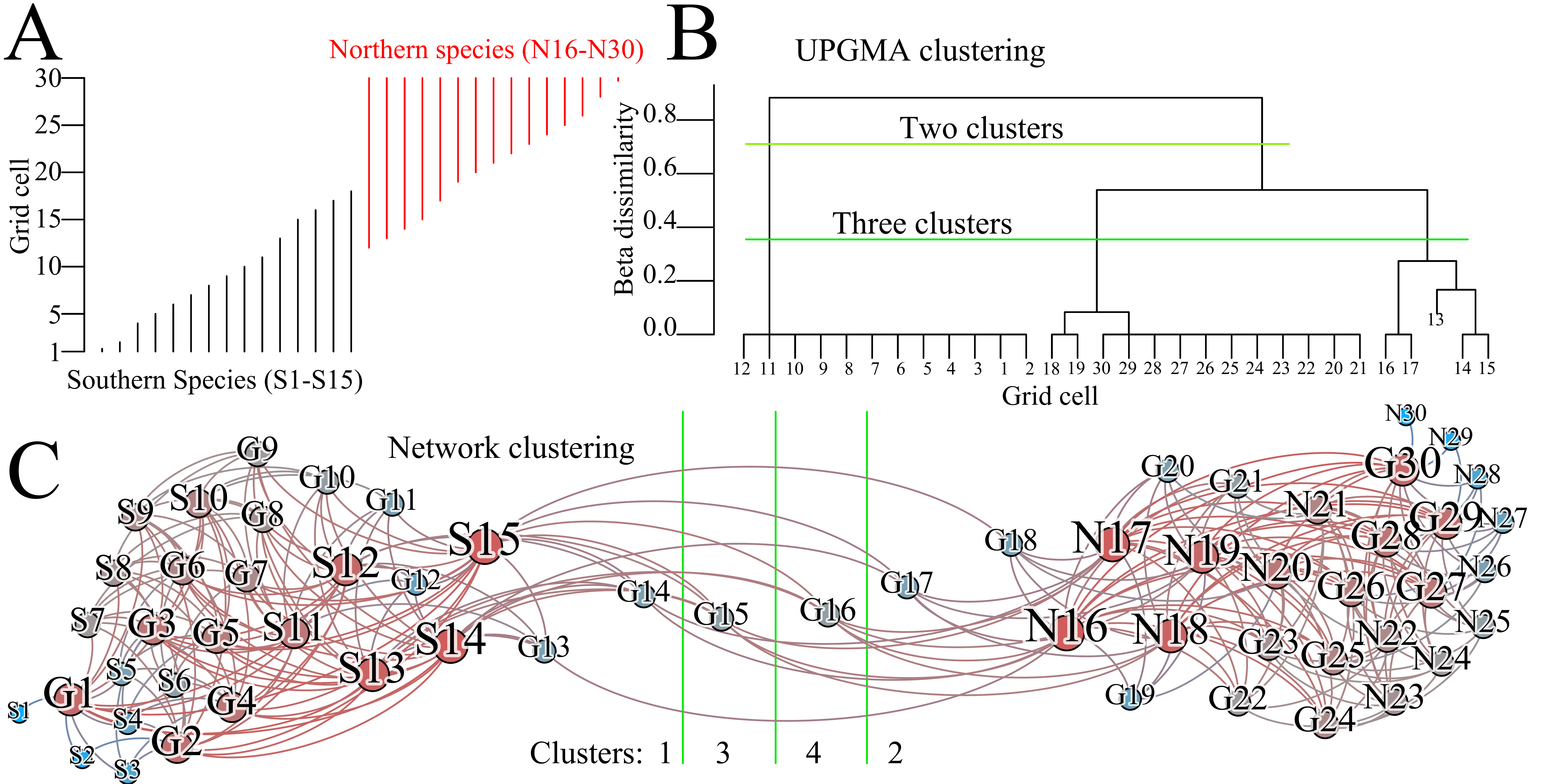}
\end{center}
\caption{
{\bf Hypothetical transition zone (N= 30 species)}. A) Species range data across a line of grid cells. The data represent two biotic assemblages that blend together in a transition zone. B) After clustering the data according to Simpson's similarity index, the best representation of the data is as two or three clusters, but three causes the transition zone to appear as a distinct Biogeographical regions. C) In the network clustering, the best representation is as two or four clusters, with four being optimal (shown). In this optimal partition scheme, the transition zone is clustered alone but without any species, revealing it as a zone that does not definitively belong to either fauna. In the two cluster solution, the grid cells are divided evenly between the faunas. Colours indicate the number of links that each node has: grid cells with higher richness and species with larger ranges are redder, while grid cells with less richness and species with smaller ranges are bluer. The sizes of the nodes are similarly proportional. ``G" denotes grid cell, ``N" denotes Northern species, and ``S" denotes southern species. Node positions were determined with the Force Atlas algorithm in the Gephi software package \cite{bastian2009gephi}.
}
\label{Figure 5}
\end{figure*}

\section{Discussion}

Our network analyses of empirical and hypothetical datasets reveal important differences as compared to approaches based on species similarity. These are not only quantitative in terms of resolution -- i.e. the total number of regions identified -- but also qualitative, affecting both the areas and the boundaries of bioregions and realms.

\textbf{Global amphibian bioregions.} The differences in number of zoogeographic realms and bioregions found by our network method as compared to the similarity analysis by Holt \textit{et al}. \cite{holt2013update} do not arise from a lower cutoff threshold for our approach, because we followed their procedure for merging regions with less than 10 grid cells into the closest regions. Rather, we interpret this difference as stemming from a fundamental difference in methodology -- our approach clusters patterns of presence-absence relationships while theirs identifies clusters of grid cells with low distributional and phylo-distributional turnover. 

Our results suggest that, at least for amphibians, turnover measures based on species distribution data alone is sufficient to identify realm boundaries. This conforms with the distribution-only approach undertaken in Holt \textit{et al}. \cite{holt2013update}, which similarly identifies Weber's line as the realm boundary between the Oriental and Oceanian faunas, although it does not identify Wallace's line. This suggests that Weber's line may be more robust and independent of methodology than Wallace's line. 

At a finer scale, our analysis was able to recover many expert based bioregion scale regions around the world, often approaching WWF's classification. Taking South America as an example, our analysis not only identified the 2--3 major regions found by Holt \textit{et al}. \cite{holt2013update}, but also successfully recovered climatically and physiognomically distinct bioregions such as the seasonally dry and fire-prone Brazilian Cerrado, the evergreen Atlantic forest of eastern Brazil, and the geologically old and nutrient-poor Guianan highlands, among several others regions that were not recognised by our benchmark example using similarity \cite{holt2013update}. We also note some important differences in the area and delimitation of these bioregions. The western limits of the Amazonian region inferred by Holt \textit{et al}. \cite{holt2013update}, for instance, cuts across the Andean mountains, despite the enormous altitudinal and physiological differences between these two regions. Our delimitations better conform to the commonly recognised boundaries between the Andes and Amazonia \cite{ter2013hyperdominance}, thus reflecting not only taxonomic differences but also current topography, climate, and evolutionary history \cite{hoorn2010amazonia}.

\textbf{United States plant bioregions.} Similarly to the analysis of amphibian data, the species clustering approach based on similarity exhibited both quantitative as well as qualitative shortcomings: it was unable to distinguish more than a few Biogeographical regions under its optimal clustering, and it was heavily biased by political state boundaries (Fig. 4). 

These shortcomings are perhaps unsurprising given a few challenges of the task, which we chose to illustrate the potential pitfalls encountered in empirical datasets of species distribution. \textit{First}, we clustered raw occurrence data as presence or absence of a species in a county. This becomes evident in the output of the similarity analyses, as presence/absence data is often compiled at the state rather than the county level, producing apparently unique floras at the state level (mostly evident in Fig. 4, right column). \textit{Second}, county sizes differ substantially, creating an artifactual richness bias that is correlated with county size (Fig. S1). This pitfall might have been avoided by aggregating data by grid cell, but it should be already minimised by the Simpson's similarity index utilised. 

\textbf{Occurrence data vs SDMs.} Predicting species ranges through Species Distribution Models (SDMs) would have been an alternative approach for bioregion inference in cases with documented geographic biases, such as for American plants \cite{morueta2013habitat}. This would have reduced the bias of identifying political state boundaries as bioregion limits in our empirical example using the similarity approach (Fig. 4). However, we identify two inherent pitfalls. \textit{First}, SDMs are still largely sensitive to the data and methodology used \cite{duputie2013wild}, carrying its own set of problems and assumptions -- such as reliance on interpolated climatic data, general unavailability of non-climatic niche variables, and exclusion of potentially crucial lineage specific traits such as dispersal ability, biotic interactions, population dynamics and evolutionary history \cite{guisan2005predicting}. \textit{Second}, using SDMs for bioregion inference could become conceptually circular. If we are to understand how species cluster into distinct bioregions, and how the boundaries of these bioregions relate to environmental gradients, this comparison needs to be post-hoc. We cannot use SDMs to delineate the same bioregions that are then used to compare their correspondence to environmental variables, because these variables are already the major component of SDMs. 

Considering these pitfalls, we argue that new methods for biogeographic delineation must be designed around the current challenges offered by real occurrence data -- which are geographically and taxonomically biased, but nevertheless constitute the most reliable evidence on where species occur. Here we have provided qualitative and quantitative evidence that a network based approach outperforms current methods based on similarity. 

\section{Methods}

To classify bioregions based on species distribution data we hierarchically classify groups of species and grid cells into realms and bioregions. To achieve this goal we borrow from the techniques developed in network science to create a network that will be meaningful for biogeographic analysis, and then use network clustering algorithms to hierarchically partition groups of nodes into clusters. In this paper we adapt the methodology presented by Vilhena \textit{et al}. \cite{vilhena2013bivalve} and Sidor \textit{et al}. \cite{sidor2013provincialization} for modelling species distributions as a network. First, we build the network we will cluster, and second, we choose the algorithm that will cluster our network to produce bioregions.

\subsection{Delimiting bioregions with networks}

A bipartite network (Fig. 2) has two disjoint sets of nodes with no links between nodes of the same set. Many biological systems have been abstracted as bipartite networks, such as plant-pollinator interactions inferred by visitation \cite{bascompte2003nested}, sexual contact between heterosexual partners \cite{ergun2002human}, and interactions between prey and bait proteins generated by yeast two-hybrid screening, an experimental method to test whether pairs of proteins interact \cite{uetz2000comprehensive}. 

The geographic relationships between species and localities can also be abstracted as a bipartite association network, where links are the occurrences of species within geographic locations. Interpretations derived from analyses of presence-absence networks are comparable with plant-pollinator networks, because relationships between entities of the same set are associational, such as co-visitation and co-occurrence. Second order relationships in presence-absence networks are paths of length two, or 2-paths. The number of 2-paths between species is the number of times those species co-occur, while the number of 2-paths between a pair of localities, regions, or grid cells is the number of species shared by both grid cells. Although second order range overlaps between two species may not be directly intuitive biologically, in practice it should allow the delimitation of bioregions comprised of only partially overlapping species. Partial occupancy of a species' potential range (Fig. 1A) may be due to intrinsic traits (e.g. dispersal ability, tolerance to specific climatic and environmental variables, ecological interactions) as well as the region's physical features (e.g. soil and climatic heterogeneity, geological history, presence of dispersal barriers). 

A more complicated pattern is the number of joint occurrences, where two species occupy the same two localities. This can be measured as the number of 4-paths that complete a loop (Fig. 2A). These relationships can be combined to reveal properties of geographic ranges. For example, the number of 3-paths between a species A and locality B divided by the number of 2-paths exiting from species A is the fraction of co-occurrences of species A that also occupy locality B. By setting up the machinery to capture ``higher-order" patterns, we can detect complex patterns of presence-absence. 

The adjacency matrix $A$ of this network formally expresses species occurrences, and is written 

\begin{equation}
A_{ij} = 
\begin{cases}
  1 & \text{if node $i$ is linked with node $j$ }  \\ 
  0 & \text{otherwise} .
\end{cases}
\end{equation}

For the rows and columns of this matrix, we order first by species $(1...n)$ and second by grid cells $(n+1...n+m)$, producing a square matrix with $n+m$ rows and $n+m$ columns. This is expressed

\begin{equation}
A = \begin{pmatrix} 0 & B \\ B^T & 0 \end{pmatrix} \, ,
\end{equation}

\noindent where $B$ is the binary presence-absence matrix, in which rows are taxa and columns are localities. The upper left block and lower right block in this matrix are zeroes, because species cannot occur in species and localities cannot occur in localities. The square of the adjacency matrix $A$ gives the co-occurrence matrix $C$ between taxa as the upper left square, or number of co-occurrences between pairs of species, and the matrix of shared species $S$ as the bottom right square, or number of shared species between pairs of grid cells

\begin{equation}
A^2 = \begin{pmatrix} C & 0 \\ 0 & S \end{pmatrix} \, ,
\end{equation}

\noindent where elements in the upper right and lower left squares of the matrix are zeroes because 2-paths are exclusively between two species or between two localities. Total paths of length $i$ between nodes can be expressed by raising the matrix to the $i$th power. By formulating the data in this new measures can be derived and tools from network theory can be readily applied. In the next section we apply a common clustering algorithm to this bipartite network.

\subsection{Clustering the bipartite species network}

Among candidate clustering algorithms, the map equation is the most suitable approach to be extended to bipartite networks \cite{rosvall2008maps,rosvall2011multilevel,sidor2013provincialization}. The map equation is a general approach that, for our purposes, corresponds to an intuitive process. First, the algorithm chooses a random grid cell. It then randomly chooses a species found in that grid cell, examines the geographic range of that species, and selects a grid cell at random within its geographic range. It repeats this process iteratively and exhaustively. In biota with substantial biogeographic structure, the algorithm would spend long time intervals within bioregions, crossing only when it selected a cross-bioregion species. 

If the algorithm would be requested to report a list of the grid cells and species chosen, it would save time to simply list the bioregions visited. The map equation quantifies the tradeoff between losing detail from all visits and saving time by communicating a shorter list; in biota with strong biogeographic structure, it will be better on average to communicate a shorter list of visits. The map equation has been extended to deal with hierarchical partitions, which we use to reveal realms and bioregions \cite{rosvall2008maps,rosvall2011multilevel}. The software packages for the two level and hierarchical approaches are available online (http://www.mapequation.org).

\subsection{Method validation and performance}

As a first empirical test case, we apply the network clustering method developed here to the International Union for Conservation of Nature (IUCN) amphibian database, which contains range shape files for each of the c. 6,100 included species. We use only native ranges for the analysis. We choose to analyse distributional data for amphibians \cite{IUCN} because \textit{i)} we consider this database to be thoroughly verified by the scientific community; \textit{ii)} we expect that the eco-physiological tolerance of the amphibians should be narrower than that for e.g. mammals or birds, and therefore more closely track bioregion level regions; and \textit{iii)} this would allow a direct comparison with a recent study by Holt \textit{et al}. \cite{holt2013update}, where both species distribution data alone and combined with phylogenetic information was used to infer zoogeographic regions and realms on a global level. 

Our second empirical test is performed using the United States Department of Agriculture (USDA) plant database, which contains the presence or absence of 22,918 native vascular plant taxa (corresponding to 17,600 species) spread through 50 states and 3143 counties of the USA. We use only the range of native plants, delineating bioregion structure for all plants, only trees, and all plants but trees (i.e. herbs, lianas, shrubs, subshrubs and vines). These data are ideal as a benchmark because they contain several challenges for computational methods. \textit{First}, United States county areas are longitudinally biased, with larger counties in the west and smaller counties in the east (Fig. S1). \textit{Second}, plant distributions are aggregated differently across states, causing systematic compositional biases across state borders. \textit{Third}, counties are unevenly sampled. To our knowledge, no quantitative bioregion delineation of these data are available for direct comparison. 

Lastly, we use a recent hypothetical dataset to illustrate key differences between our network method and species similarity approaches. In a recent commentary by Kreft \& Jetz \cite{holt2013updatereply}, this dataset was used to showcase potential pitfalls for selecting the wrong number of clusters. The hypothetical data contains a transition zone, where the most widespread species in a Northern and Southern biota co-occur (Fig. 5A). In their analysis \cite{holt2013updatereply}, the number of clusters selected as optimal was shown to fully determine whether or not the transition zone appeared as a distinct Biogeographical regions. This result was used to illustrate the danger of classifying transition zones as distinct Biogeographical regions, but also highlights the sensitivity of inferring bioregions based on species similarity measures. 

To assess the performance of our network based clustering with a conventional species similarity approach, we opted for the methodology selected as best in a recent methods review by Kreft \& Jetz \cite{kreft2010framework}. To apply that approach to our plant data, we created a matrix of counties and computed the species similarity between each pair of US counties with species data. We applied the $\beta_{sim}$ index to the different datasets, written $1-\frac{a}{min(b,c)+a}$. Here $a$ is the number of shared species between two species assemblages and $b$ and $c$ are the total unique species to either assemblage (quadrat, locality, grid cell, etc). Note that $\beta_{sim}$ is 0 when the species assemblages are either identical or the smaller assemblage is a subset of the larger assemblage, and $\beta_{sim}$ is 1 when the assemblages contain no shared species. This similarity index was originally proposed by Simpson \cite{simpson1943mammals}, but it has received different uses and modifications in the literature \cite{tuomisto2010diversity}. This measure is considered ideal over more conventional measures (such as the Jaccard) because it is less sensitive to differences in species richness \cite{kreft2010framework}. 

We clustered this matrix with the Unweighted Pair Group Method with Arithmetic Mean (UPGMA) approach to generate a hierarchical dendrogram that summarises the distances between counties. From this dendrogram, we selected an optimum number of clusters by finding the ``knee" in the evaluation curve \cite{salvador2004determining}, with the average percentage of endemics as our evaluation measure \cite{kreft2010framework}.

\section{CONCLUSIONS}

Bioregion conservatism has been suggested as a crucial feature shaping the uneven distribution of the world's biota \cite{donoghue2008phylogenetic,wiens2010niche,crisp2006biome}, including the establishment and maintenance of the tropical gradient in species richness. The evolution of entire bioregions is also gaining focus in macroecological meta-analyses using phylogenetic, palaeontological and distribution data \cite{crisp2009phylogenetic,hoorn2010amazonia,holt2013update}. 

A critical component of bioregionalisations is the robust identification, delimitation and nomenclature of global bioregions. This has been limited by our use of similarity approaches that are unable to objectively identify the optimal number of clusterings that represent the data, and to infer realistic and reproducible boundaries between bioregions. Our study demonstrates that species distribution data holds alone a large and unrealised potential towards those goals. 

Phylogenetic turnover measures have been used as alternative to \cite{rosauer2014phylogenetic}, as well as in combination with \cite{holt2013update}, species distribution data. However, they rely on robust and well-sampled species level phylogenies (which are currently lacking for many organismal groups) and may introduce circularity when using the identified bioregions for measuring the degree of phylogenetic niche conservatism (as shifts in bioregions are commonly associated with speciation events).  Phylogenies -- especially when time calibrated -- can subsequently be used to shed light on the temporal origin, evolution, and phylogenetic relatedness of bioregions. 

Important challenges, however, remain in order to further advance bioregionalisations: 

\begin{enumerate}

\item \textit{Quantity and quality of species occurrence data.} Mapping the distribution of the world's estimated 8.7 million species \cite{mora2011many} constitutes a major challenge in biological research \cite{guralnick2009biodiversity} and is paramount for bioregion delineation. The ever increasing digitisation of natural history collections worldwide now offers access to over 418 million records at the Global Biodiversity Information Facility (www.gbif.org), but this figure is still far from the estimated total of one billion specimens. It is clear that the occurrence data currently available contain substantial spatial, taxonomic and temporal biases \cite{boakes2010distorted}, besides a certain proportion of errors (e.g. misidentified specimens and poorly or wrongly annotated locality information). Substantial efforts are required to revise such raw occurrence data and combine them with field observations and expert knowledge, for producing GIS based polygons of species distribution ranges (e.g., IUCN, www.mappinglife.org). 

\item \textit{Methodological development and integration.} Biomics will greatly profit from bringing together different techniques, data and disciplines. These should minimally include remote sensing, climatic mapping and bioregion modelling based on key species \cite{sarkinen2011forgotten}. New methodologies for bioregion delineation need to be reproducible and transparent about their assumptions. They should offer measures of reliability regarding the number and boundaries of species clusters identified, or parts thereof -- e.g., the delimitation of the same bioregion may be more or less robust along different edges. Finally, they should be regularly validated through ground truthing. 

\item \textit{Theory vs reality.} Are there really bioregions, how were they formed, how are they maintained through time? We still lack an elementary ecological theory for addressing these questions, despite the fact that few people contest the existence of bioregions. We also need to understand how extrinsic (e.g. climate, geological history, soils) and intrinsic (e.g. functional traits, biotic interactions, physiology) variables interplay to produce the differences we observe in the number and delimitation of bioregions based on data from plants, birds, amphibians, and mammals \citep[and this study]{holt2013update} -- and expand our inferences to many other understudied groups. 

\end{enumerate}

More than a century after the first Biogeographical regions were proposed \cite{wallace1876geographical}, we may now have enough data to delimit the world's realms and bioregions in greater detail than Wallace could ever envision. Our study however illustrates that new methodologies play a crucial role in this process, and that network methods offer a new set of exciting tools to classify, delimit and better understand biodiversity. 

\bibliography{Vilhena_Antonelli_biomes}

\section{Acknowledgements}
We thank J. Grummer for discussions on amphibian distributional patterns. We thank C.T. Bergstrom, M. Rosvall, and F. Meacham for discussions of network clustering and H. Tuomisto for help with species similarity indices. A.A. is supported by grants from the Swedish Research Council (B0569601) and the European Research Council under the European Union's Seventh Framework Programme (FP/2007--2013, ERC Grant Agreement n. 331024). 

\section{Author contributions}
D.A.V. and A.A. designed the study and wrote the paper. D.A.V. developed the method, wrote the code, and performed the analyses. 

\section{Additional information}


\textbf{FIG. S1} United States states (thicker lines) and counties (thinner lines), shown to illustrate the west to east tendency of increasing county numbers (and decreasing county areas). Adapted from Wikimedia.


\textbf{Competing financial interests:} The authors declare no competing financial interests.

\end{document}